\documentstyle[11pt]{article}
\renewcommand{\baselinestretch}{1.75} 
\def\F{{\cal F}}
\def\N{N}
\def\M{m}
\def\en{e^{(n)}}
\def\an{a^{(n)}}
\def\anm{a^{(n-1)}}
\def\anp{a^{(n+1)}}
\def\Sn{S^{(n)}}
\def\Snm{S^{(n-1)}}
\def\Snp{S^{(n+1)}}
\def\An{A^{(n)}}
\def\Bn{B^{(n)}}
\def\eps{\epsilon}
\def\cl{{\rm cl}}
\def\oneloop{{\rm 1-loop}}
\def\oneinst{{\rm 1-inst}}
\def\dinst{{d-\rm{inst}}}
\newcommand{\SU}{ {\rm SU} }
\newcommand{\U}{ {\rm U} }
\newcommand{\ttil}{ {\tilde t} }

\newcommand{\th}{\theta}
\newcommand{\la}{\lambda}
\newcommand{\La}{\Lambda}

\newcommand{\be}{\begin{eqnarray}}
\newcommand{\ee}{\end{eqnarray}}

\newcommand{\pr}{\partial}

\newcommand{\np}{\newpage}
\newcommand{\hs}{\hspace}
\newcommand{\vs}{\vspace}

\newcommand{\nn}{\nonumber}

\begin{document}

\thispagestyle{empty}

\vs*{-25mm}
\begin{flushright}
BRX-TH-446\\[-.2in]
BOW-PH-113\\[-.2in]
HUTP-98/A084\\
hep-th/9901124
\end{flushright}

\begin{center}
{\LARGE{\bf 
One-instanton predictions of Seiberg-Witten curves for product groups
}} \\
\vspace{.2in}

\renewcommand{\baselinestretch}{1}
\small
\normalsize
Isabel P. Ennes\footnote{Research supported 
by the DOE under grant DE--FG02--92ER40706.}\\
Martin Fisher School of Physics\\
Brandeis University, Waltham, MA 02454

\vspace{.1in}
Stephen G. Naculich\footnote{Research supported 
in part by the National Science Foundation under grant no. PHY94-07194.}\\
Department of Physics\\
Bowdoin College, Brunswick, ME 04011

\vspace{.1in}

Henric Rhedin\\
Martin Fisher School of Physics\\
Brandeis University, Waltham, MA 02454\\
and\\
Department of Engineering Sciences\footnote{Permanent address},\\
 Physics and Mathematics\\
Karlstad University, S-651 88 Karlstad, Sweden\\
\vspace{.1in}

Howard J. Schnitzer\footnote{Research supported in part
by the DOE under grant DE--FG02--92ER40706.\\
{\tt \phantom{aaa} naculich@bowdoin.edu; henric.rhedin@kau.se;
ennes,schnitzer@binah.cc.brandeis.edu}}\\
Martin Fisher School of Physics\footnotemark[3]\\
Brandeis University, Waltham, MA 02454\\
and\\
Lyman Laboratory of Physics\\
Harvard University, Cambridge, MA 02138\\

\vspace{.2in}

{\bf{Abstract}} \end{center}
\renewcommand{\baselinestretch}{1.75}
\small
\normalsize
\begin{quotation}
\baselineskip14pt
\noindent 
One-instanton predictions for the prepotential are obtained from the
Seiberg-Witten curve for the Coulomb branch of ${\cal N}=2$ 
supersymmetric gauge theory for the product group
$\prod_{n=1}^m \SU(N_n)$ with a massless matter hypermultiplet
in the bifundamental representation $(N_n,\bar N_{n+1})$ of
$\SU(N_n)\times \SU(N_{n+1})$ for
$n=1\,\,{\rm to}\,\,m-1$, together with $N_0$ and $N_{m+1}$
matter hypermultiplets in the fundamental representations 
of $\SU(N_1)$ and $\SU(N_m)$ respectively. 
The derivation uses
a generalization of the systematic perturbation expansion 
about a hyperelliptic curve developed by us in earlier
work. 
\end{quotation}

\np 

\setcounter{page}{1}

Spectacular advances have been made in our understanding of 
the non-perturbative
behavior of supersymmetric gauge theories and string theories. 
In particular, the program of Seiberg and Witten \cite{SeibergWitten}
allows one to compute the exact behavior of low-energy four-dimensional 
${\cal N}=2$ supersymmetric gauge theories in various regions of moduli space
from the following input: 
a Riemann surface or algebraic curve appropriate to the specific theory,
and the Seiberg-Witten meromorphic one-form. 
When the curve in question is non-hyperelliptic, however,
the explicit extraction of this information is a challenging technical problem.
In this letter,
we will use Seiberg-Witten theory to calculate the
one-instanton predictions of an ${\cal N}=2$ supersymmetric
gauge theory based on the product group $\prod_{n=1}^\M \SU (\N_n)$
by extending the methods of 
ref.~\cite{DHokerKricheverPhong1}-\cite{EnnesNaculichRhedinSchnitzer2}.

The exact low-energy properties of ${\cal N}=2$ theories 
are encapsulated in the form of the prepotential $\F (A)$,
in terms of which the Wilson effective Lagrangian is
\be
{\cal L}=\frac{1}{4\pi}{\rm Im}\left[\int{\rm d}^4\th
\frac{\pr{\cal F}(A)}{\pr A^i}\bar{A}^i+
\frac{1}{2}\int{\rm d}^2\th \frac{\pr^2{\cal F}(A)}{\pr A^i\pr A^j}
W^iW^j\right],
\label{eqnaa}
\ee
to lowest order in the momentum expansion,
where $A^i$ are ${\cal N}=1$ chiral superfields
and $W^i$ are ${\cal N}=1$ vector superfields. 
Holomorphy implies that the prepotential in the Coulomb phase 
has the form of an instanton expansion 
\be
{\cal F}(A)={\cal F}_{\cl}(A)\,+\,{\cal F}_{\oneloop}(A)\,+\,
\sum_{d=1}^{\infty} {\cal F}_{\dinst}(A).
\label{eqnaaa}
\ee

Consider an ${\cal N}=2$ supersymmetric gauge theory 
based on the gauge group $\prod_{n=1}^\M \SU(\N_n)$.
In addition to the chiral gauge multiplet in the adjoint representation
of each of the factor groups,
the theory we are considering contains a massless matter hypermultiplet in the 
bifundamental representation $(\N_n, \bar \N_{n+1})$ 
of $\SU (\N_n) \times \SU (\N_{n+1})$  for $n=1, \ldots, \M-1$;
$\N_0$ matter hypermultiplets 
in the  fundamental representation of SU$(\N_1)$
(whose masses we denote $e^{(0)}_k, 1 \leq k \leq \N_0$);
and $\N_{\M+1}$ matter hypermultiplets 
in the fundamental representation of SU$(\N_\M)$
(whose masses we denote $e^{(\M+1)}_k, 1 \leq k \leq \N_{\M+1}$).
The adjoint multiplets contain complex scalar fields $\phi^{(n)}$
for each of the factor groups. 
The Lagrangian has a potential with  flat directions, along which 
the symmetry is generically broken to $\prod_{n=1}^\M \U (1)^{\N_n-1}$. 
The moduli space of the theory is therefore  parametrized by
the order parameters
$\en_k$ 
($1\leq k\leq N_n$, $1\leq n \leq \M$), 
which are the eigenvalues of the $\phi^{(n)}$
and satisfy the constraint $\sum_{k=1}^{N_n} \en_k = 0$.

The curve for this product group theory was obtained 
using M-theory \cite{Witten}
and geometric engineering \cite{Katz}, 
and made more explicit in ref. \cite{ErlichNaqviRandall}:
\be
  P_{\N_0} (x) \,t^{\M+1} \,
-\, P_{\N_1} (x) \,t^\M\,     
+ \,\sum_{j=0}^{\M-1} (-)^{\M-j+1}
  \left[ \prod_{\ell=1}^{\M-j} L_\ell^{\M-\ell-j+1} \right] 
  P_{\N_{\M-j+1}}(x)\, t^j\, 
=\,0,
\label{eqna}
\ee
where 
\be
P_{\N_n} (x)  &=&  \prod_{i=1}^{\N_n} \,(x-\en_i) ,
\hs{.4in}(n=0 \,\,{\rm to}\,\,m),\cr\cr
L_n^2        &=&  \La_n^{2\N_n -\N_{n-1} - \N_{n+1} }, 
\label{eqnb}
\ee
with $\La_n$ the quantum scale of the gauge group $\SU (\N_n)$. 
The requirement of asymptotic freedom, 
and restriction to the Coulomb phase, 
implies that $\La_n$ appear with positive powers in (\ref{eqna}). 

The curve (\ref{eqna}) 
describes a $(\M+1)$-fold branched covering 
of the Riemann sphere, 
with sheets $n$ and $n+1$ connected by $\N_n$ 
square-root branch-cuts centered about $x=\en_k$
($k=1$ to $N_n$),
and having endpoints $x_k^{(n)-}$  
and $x_k^{(n)+}$.
Following the approach of  Seiberg-Witten,
we will use this curve to compute the
renormalized order parameters and their duals
\be
2\pi i \an_k
=\oint_{\An_k} \la 
\hs{10mm} {\rm and} \hs{10mm}
2\pi i \an_{D,k}
=\oint_{\Bn_k}\la,
\label{eqnbb}
\ee
where $\la$ is the Seiberg-Witten differential,
and  $\An_k$ and $\Bn_k$  ($2 \leq k \leq \N_n$)
are a set of canonical homology cycles for the Riemann 
surface.
The cycle $\An_k$ is chosen to be a simple contour on sheet $n$
enclosing the branch cut centered about $\en_k$.
The cycle $\Bn_k$ goes from 
$x_1^{(n)-}$ to $x_k^{(n)-}$ on the $n^{\rm  th}$ sheet and from 
$x_k^{(n)-}$ to $x_1^{(n)-}$ on the $(n+1)^{\rm th}$
\cite{DHokerKricheverPhong1}.
Once we  obtain $ \an_k$ and $ \an_{D,k} $,
the prepotential can be computed by integrating 
\be
\an_{D,k}=\frac{\pr {\cal F}}{\pr \an_k}\,,  
\hs{15mm} (k=2\,\, {\rm to}\,\,\N_n, \hs{5mm} n=1\,\, {\rm to}\,\,\M).
\label{one}
\ee

In our computation, 
we  will perform a multiple perturbation expansion 
in the several quantum scales $L_n$,
with  the result for the prepotential 
 given to one-instanton accuracy, 
{\it i.e.}, $O(L_n^2)$ for all $n$.
To calculate $\an_k$ and $\an_{D,k}$ for the group $\SU (\N_n)$,
we define
$ t = \ttil \prod_{\ell=1}^{n-1} L_\ell^2 $
to recast the curve (\ref{eqna}) as 
\be
 \sum_{j=0}^{n} (-)^j 
  \left[ \prod_{\ell=j+1}^{n-1} L_\ell^{2(\ell-j)} \right]
  P_{\N_j} (x) \,\ttil^{\M+1-j} \,
+ \,\sum_{j=n+1}^{\M+1} (-)^j 
  \left[ \prod_{\ell=n}^{j-1} L_\ell^{2(j-\ell)} \right]
  P_{\N_j} (x)\, \ttil^{\M+1-j} = 0, 
\label{eqnc}
\ee
or more explicitly
\be
\cdots &+& 
  	  (-)^{n-3} L_{n-2}^2 L_{n-1}^4    P_{\N_{n-3}} (x) \,\ttil^{\M-n+4}
\,\,+\,\, (-)^{n-2} L_{n-1}^2		   P_{\N_{n-2}} (x)\, \ttil^{\M-n+3}\cr
   &+&    (-)^{n-1}           		   P_{\N_{n-1}} (x) \,\ttil^{\M-n+2} 
\,\,+\,\, (-)^{n  }           	           P_{\N_{n  }} (x) \,\ttil^{\M-n+1} 
\,\,+\,\, (-)^{n+1} L_{n  }^2              P_{\N_{n+1}} (x) \,\ttil^{\M-n  }\cr
   &+&    (-)^{n+2} L_n^4 L_{n+1}^2        P_{\N_{n+2}} (x) \,\ttil^{\M-n-1} 
\,\,+\,\, (-)^{n+3} L_n^6 L_{n+1}^4 L_n^2  P_{\N_{n+3}} (x) \,\ttil^{\M-n-2} 
\,\,+\,\,   \cdots \,\, = \,\, 0. \,\,\,\,\,\,\,\,\,\,\,
\label{eqnd}
\ee

To obtain the one-loop, zero-instanton contribution,
{\it i.e.}, $O(\log L_n)$, to $\an_k$ and $\an_{D,k}$,
one may set $L_\ell =0$ for $\ell \neq n$,
in which case the curve (\ref{eqnd}) reduces,
after the change of variable $  \ttil = y/P_{\N_{n-1}}(x) $,
to the hyperelliptic curve
\be
y^2 \,+\, 2 A(x) y\, +\, B(x)  \,=\, 0,
\label{eqndd}
\ee
with
\be
A(x) &=& - {\textstyle {1\over 2}} P_{\N_n} (x), \cr
B(x) &=& L_n^2 P_{\N_{n+1}} (x) P_{\N_{n-1}} (x) .
\label{eqne}
\ee
On one of the sheets, eq.~(\ref{eqndd}) has the solution
\be
y = -A-r   \hskip.5in {\rm where} \hskip.2in    r=\sqrt{A^2-B},
\label{eqnf}
\ee
from which we may compute the Seiberg-Witten differential 
$ \la =x  {\rm d}y/{y} $
in the hyperelliptic approximation to be
\be
\lambda_{I} \,=\,   
{ x\left( {A'\over A}-{B'\over {2B}}\right)
\over{\sqrt{1-{B\over A^2}}}}
{\rm d}x.
\label{eqnh}
\ee
(On the other sheet, the solution to eq.~(\ref{eqndd}) is $y=-A+r$).

To obtain the one-instanton correction 
({\it i.e.},  $O(L^2_\ell)$ for all $\ell\,$)
to the order parameters,
one again makes the change of variables 
$\,\tilde t\,=\,y/P_{N_{n-1}}(x) \,$ in eq.~(\ref{eqnd}),
and keeps two more terms beyond those in eq.~(\ref{eqndd}), 
obtaining the quartic curve
\be
\eps_1 (x) y^4 \,+\, y^3\, +\, 2 A(x) y^2\, +\, B(x) y\, +\, \eps_2 (x) \,=\,0, 
\label{eqni}
\ee
where
\be
\eps_1 (x) &=& - L_{n-1}^2 \,{P_{\N_{n-2}}(x) \over  P_{\N_{n-1}}^2 (x) },\cr\cr
\eps_2 (x) &=& - L_{n}^4\, L_{n+1}^2 \,P_{\N_{n+2}} (x)\, P_{\N_{n-1}}^2 (x).
\ee
Rewriting eq.~(\ref{eqni}) as
\be
y\,+\,A\,+\,r\,=\, {1 \over y+A-r}
\left[ - \eps_1 y^3 \,-\, {\eps_2 \over y} \right]
\ee
and substituting $y=-A-r$ into the right hand side, 
which is already first order in $\eps$, 
we obtain
\be
y = -A\,-\,r\,-{(A+r)^3\over 2r} \eps_1 \,-\, {1 \over 2r (A+r)} \eps_2
\,+\,\cdots,
\ee
to first order in $\eps_1$ and $\eps_2$. 
The Seiberg-Witten differential is correspondingly modified to
\be
\lambda \,=\, \lambda_{I}\,+\lambda_{II}\,+\,\cdots ,
\label{eqnj}
\ee
where $\lambda_I$ is the hyperelliptic approximation (\ref{eqnh}) and 
\be
\lambda_{II}   =
\left( -A\, \eps_1\, -\, {A \over B^2} \,\eps_2 \right)
{\rm d}x 
= - L_{n-1}^2 \,{P_{\N_n}  P_{\N_{n-2}} \over 2 P_{\N_{n-1}}^2  }\, {\rm d}x 
  - L_{n+1}^2 \,{P_{\N_n}  P_{\N_{n+2}} \over 2 P_{\N_{n+1}}^2  }\, {\rm d}x, 
\ee
obtained from a calculation similar to that in Appendix C of 
ref.~\cite{NaculichRhedinSchnitzer}.

One computes the order parameters (\ref{eqnbb})
using the methods of 
refs.~\cite{DHokerKricheverPhong1,NaculichRhedinSchnitzer,
EnnesNaculichRhedinSchnitzer},
obtaining  
\be
\an_k \,=\, \en_k\,+
\,{1 \over 4} \,L_n^2 \,{\pr \Sn_k\over \pr x} (\en_k) + \cdots\,,
\hs{15mm} (k=1\,\, {\rm to}\,\,\N_n, \hs{5mm} n=1\,\, {\rm to}\,\,\M),
\ee
where the residue functions $\Sn_k(x)$ are defined in terms of eq. 
(\ref{eqne})  by
\be
{ \Sn_k (x) \over (x - \en_k)^2  } =  { B(x) \over A(x)^2 }\,.
\ee
Considerations analogous to those of Appendix D of ref. 
\cite{NaculichRhedinSchnitzer} give the identities
\be
\sum_{j=1}^{\N_n} \,{\pr \Sn_j \over  \pr x}\, (\en_j)\,=\,0,
\label{eqnk}
\ee
implying $ \sum_{i=1}^{\N_n} \,\an_i = \sum_{i=1}^{\N_n} \,\en_i $
to the order that we are working. 

Next, the dual order parameters $\an_{D,k}$
are computed along the lines of sec. 5 of ref. \cite{NaculichRhedinSchnitzer},  
giving
\be
2\pi i\,\an_{D,k} 
&=&
\,[2\N_n-\N_{n+1}-\N_{n-1}+2\,{\rm log}\,L_n+{\rm const}]
\,\an_k 
\,-\, 2\,\sum_{i\neq k}^{\N_n} (\an_k- \an_i)\,\log\,(\an_k-\an_i)\,
\nn \\
 &+&  \sum_{i=1}^{\N_{n+1} }(\an_k- \anp_i)\,\log\,(\an_k-\anp_i) 
\,\,+\,\, \sum_{i=1}^{\N_{n-1} }(\an_k- \anm_i)\,\log\,(\an_k-\anm_i) \nn \\
&+&
  {1 \over 4} L_n^2\, {\pr \Sn_k \over \pr x}\,(\an_k)
\,\,-\,\, {1 \over 2} L_n^2\, \sum_{i\neq k}^{\N_n} 
		{\Sn_i (\an_i)\over \an_k-\an_i}
\,\,+\,\, {1 \over 4} L_{n+1}^2 \,\sum_{i=1}^{\N_{n+1} }
               	{\Snp_i (\anp_i) \over \an_k-\anp_i} \nn \\
&+& {1 \over 4} L_{n-1}^2 \,\sum_{i=1}^{\N_{n-1} }
               {\Snm_i (\anm_i) \over \an_k-\anm_i} 
\,\,+\,\, \cdots ,
\hs{15mm} (k=2\,\, {\rm to}\,\,\N_n, \hs{5mm} n=1\,\, {\rm to}\,\,\M).
\label{eqnl}
\ee
In eq.~(\ref{eqnl}),
we define $a^{(0)}_k = e^{(0)}_k$
and    $a^{(\M+1)}_k = e^{(\M+1)}_k$
(the masses of the hypermultiplets in the 
fundamentals of  SU$(\N_1)$ and SU$(\N_\M)$ respectively),
and $L_0= L_{\M+1} = 0$.

One then integrates eq.~(\ref{one}) using eq.~(\ref{eqnl}) 
to obtain the prepotential (\ref{eqnaaa}) to one-instanton accuracy, 
finding
\be
{\cal F}_{\oneloop}
& =  & {i\over 8\pi} \sum_{n=1}^{\M}
\sum_{i,j=1}^{\N_n}\,
(\an_i-\an_j)^2  \log  (\an_i-\an_j)^2 \nn \\
& - &  {i\over 8\pi} \sum_{n=0}^{\M} 
\sum_{i=1}^{\N_n}\,\sum_{j=1}^{\N_{n+1}}\,
(\an_i-\anp_j)^2  \,    \log (\an_i-\anp_j)^2,     
\label{eqnm}
\ee
and
\be
{\cal  F}_{\oneinst}\,=
\,{1\over 8 \pi i} \,\sum_{n=1}^\M  \, L_n^2\, 
\sum_{k=1}^{\N_n} \,\Sn_k\,(\an_k)\,,
\label{eqnn}
\ee
where 
\be
\Sn_k (x) =
{4  \, \prod_{i=1}^{\N_{n+1}} \,(x-\anp_i) \,
     \prod_{i=1}^{\N_{n-1}} \,(x-\anm_i) \over
     \prod_{i\neq k}^{\N_n}\, (x-\an_i)^2} ,
\hs{15mm} (k=1\,\, {\rm to}\,\,\N_n, \hs{5mm} n=1\,\, {\rm to}\,\,\M).
\,\,\,\,\,\label{eqnp}
\ee
Note that as $\Sn_k (\an_k)$ depends on $\anp_i$ and  $\anm_i$
as well as $\an_i$,
eq. (\ref{eqnn}) is {\it not} just the 
naive sum of instanton contributions from each subgroup. 

The one-loop prepotential (\ref{eqnm}) 
agrees with the perturbation theory result 
for a chiral gauge multiplet in the adjoint representation
of each of the factor groups,
a massless matter hypermultiplet in the 
bifundamental representation $(\N_n, \bar \N_{n+1})$ 
of $\SU (\N_n) \times \SU (\N_{n+1})$  for $n=1$ to $\M-1$,
$\N_0$ matter hypermultiplets with masses $a^{(0)}_k$
in the  fundamental representation of SU$(\N_1)$, and 
$\N_{\M+1}$ matter hypermultiplets with masses $a^{(\M+1)}_k$
in the fundamental representation of SU$(\N_\M)$.

One check of the one-instanton correction (\ref{eqnn}) 
is provided by ref.~\cite{DHokerKricheverPhong3}, 
where various decoupling limits for ${\cal N}=2$ $\SU(N)$ gauge theory 
with a massive hypermultiplet in the adjoint
representation are considered.  
D'Hoker and Phong \cite{DHokerKricheverPhong3} obtain 
${\cal F}_{\oneinst}$  for the product group theory, 
but with restriction to a single quantum scale. 
We find agreement with their result
when we restrict eqs.~(\ref{eqnn}) and (\ref{eqnp}) 
to the special case of a single quantum scale, 
which therefore provides a test of the curve (\ref{eqna})
obtained from M-theory. 

In this paper, we showed that 
to compute the order parameters of the 
${\cal N}=2$ gauge theory for the product group $\prod_{n=1}^\M \SU(N_n)$ 
to one-instanton accuracy, 
one need only consider the sequence of quartic curves (\ref{eqni}), 
even though the complete curve for the theory (\ref{eqna}) 
is of higher order ({\it viz.}, $m+1$) for $m > 3$, 
{\it i.e.}, for products of three or more groups.
(The case $m=2$ was analyzed in ref.~\cite{EnnesNaculichRhedinSchnitzer2}.)
In the language of type IIA string theory, 
this means one need only consider all possible chains 
of four parallel neighboring NS 5-branes, 
among the total set of $m+1$ parallel NS 5-branes, 
to achieve one-instanton accuracy. 
For higher instanton accuracy, additional parallel 5-branes are required. 
An analogue of this result
plays a crucial role in our analysis of the prepotential 
and Seiberg-Witten curve for $\SU(N)$ gauge theory
with two antisymmetric and 
$N_f$ fundamental hypermultiplets \cite{EnnesNaculichRhedinSchnitzer3}.  

\vs{2mm}

\noindent{\bf{Acknowledgement:}} HJS wishes to thank the Physics Department 
of Harvard University for their continued hospitality, 
and to the CERN theory group for hospitality during summer 1998.

\vs{10mm}

\end{document}